\documentclass[%
 aip,
 jmp,%
 amsmath,amssymb,
 reprint,%
]{revtex4-1}

\usepackage{graphicx}
\usepackage{dcolumn}
\usepackage{bm}
\usepackage{hyperref}
\usepackage{multirow}
\usepackage{array}
\usepackage{booktabs}
\newcolumntype{M}[1]{>{\centering\arraybackslash}m{#1}}
\usepackage{textgreek}

\begin{document}

\title{Metadynamics for Training Neural Network Model Chemistries: a Competitive Assessment.}

\author{John Herr}
\author{Kun Yao}
\author{Ryker McIntyre}
\author{David W. Toth}
\author{John Parkhill}%
 \email{jparkhil@nd.edu}
\affiliation{%
 Dept. Of Chemistry and Biochemistry, The University of Notre Dame du Lac 
}%

\date{\today}
\begin{abstract}
Neural network (NN) model chemistries (MCs) promise to facilitate the accurate exploration of chemical space and simulation of large reactive systems. One important path to improving these models is to add layers of physical detail, especially long-range forces. At short range, however, these models are data driven and data limited. Little is systematically known about how data should be sampled, and `test data' chosen randomly from some sampling techniques can provide poor information about generality. If the sampling method is narrow `test error' can appear encouragingly tiny while the model fails catastrophically elsewhere. In this manuscript we competitively evaluate two common sampling methods: molecular dynamics (MD), normal-mode sampling (NMS) and one uncommon alternative, Metadynamics (MetaMD), for preparing training geometries. We show that MD is an inefficient sampling method in the sense that additional samples do not improve generality. We also show MetaMD is easily implemented in any NNMC software package with cost that scales linearly with the number of atoms in a sample molecule. MetaMD is a black-box way to ensure samples always reach out to new regions of chemical space, while remaining relevant to chemistry near $k_bT$. It is one cheap tool to address the issue of generalization.  
\end{abstract}

\maketitle

\section{Introduction}
\indent Neural network model chemistries (NNMCs) show incredible promise as a black-box method to transfer the accuracy of \emph{ab-initio} techniques onto large systems with many orders-of-magnitude less computational cost. They can be used to calculate energies and other molecular properties.\cite{rupp2012fast,hansen2015machine,lopez2014modeling,pilania2013accelerating,schutt2014represent,ma2015machine,janet2017resolving,janet2017predicting} Diverse network types have been introduced to model energies as a sum of atoms \cite{behler2007generalized,behler2011atomcentered,gastegger2017machine,smith2017ani} or as a sum of bonds,\cite{yao2017intrinsic} a many-body expansion,\cite{yao2017many} and even generalized hierarchical schemes.\cite{lubbers2017hierarchical,huang2017chemical} Hybrid approaches which add long-range physics are another topic of growing interest.\cite{yao2017tensormol,morawietz2016van,artrith2011high,ho2016ab} In the direction of further predictive power at higher cost, neural networks are being used to augment electronic structure theory.\cite{yao2016kinetic,brockherde2017bypassing,snyder2013orbital,snyder2012finding,li2016understanding,vu2015understanding} Tools developed by these techniques are being applied to produce new drug candidates and materials.\cite{li2017learning,ramsundar2017multitask,gomes2017atomic,ramsundar2015massively,olivares2011accelerated,hachmann2011harvard,hachmann2014lead,isayev2015materials} The diversity of the field is growing more rapidly than we can fully review in the introduction of this paper.\cite{huo2017unified,bereau2017non,rupp2015machine,behler2011neural,shakouri2017accurate,behler2017first,han2017deep,khaliullin2011nucleation,medders2013critical,medders2015representation,moberg2017molecular,riera2017toward,reddy2016accuracy,manzhos2006random,manzhos2009fitting,manzhos2014neural,deringer2017machine} However, all of these models are data-driven, and so they all must somehow choose a representative set of data. This paper assesses the performance of methods for sampling geometries. We show that because NNMCs operate in a regime of overfitting, sampling is critical and ordinary MD is inadequate for generating training data. We describe a variant of MetaMD, which provides results comparable to NMS. 

\indent Data-driven model chemistries rely on `representative sampling' of the high-dimensional space that atoms explore. To obtain a transferable model, different molecular bonding patterns and points in the potential energy surface (PES) of a stable molecule must be explored. Exhaustive sampling is impossible, so practitioners employ reasonable ideas borrowed from statistical mechanics (NVT molecular dynamics\cite{gastegger2017machine,chmiela2017machine,schutt2017quantum,grisafi2017symmetry}) and traditional force-field development (Normal Mode Sampling\cite{smith2017ani,rupp2015machine}) to provide data. Relatively little systematic knowledge exists comparing different geometry sampling methods, but it is of critical importance because the generation of samples is the limiting factor for producing a NNMC. In this paper we show that MetaMD\cite{laio2002escaping} is  useful for the purposes of sampling geometries for Neural-Network force fields, and we compare it to reasonable alternatives. We also discuss how MetaMD can be easily implemented in machine learning frameworks used to produce NNMCs.

\indent This work draws a boundary between the problem of accurately learning the PES of a molecule and obtaining transferable models for all molecules. The latter problem involves choosing bond-connectivity that is representative of molecules chemists care about. Other authors have proposed solutions to the problem of choosing molecules including genetic algorithms\cite{browning2017genetic} or building up molecules from fragments\cite{huang2017chemical} to increase transferability. One can also choose molecules from databases of reported structures. This work assumes that some ensemble of molecules has been chosen for which an accurate PES is desired.

\indent One important goal for a NNMC is to provide an accurate force $-\nabla V(\vec{R})$, for all $\vec{R}$ which might be visited in molecular dynamics trajectories and/or geometry optimization. Naturally one approach to sampling these geometries is canonical MD, but this approach has two disadvantages. The first is the high degree of repetitive sampling. On the period of typical bonds, nearly identical geometries will be sampled repeatedly. The second issue is that high energy configurations are visited with exponential infrequency. This biases neural networks towards providing higher accuracy in regions of configuration space with dense sampling, in turn hindering the network's accuracy in high-energy undersampled regions. To increase the spread of configuration space sampled during an MD trajectory, one may increase the temperature to increase the exploration of high-energy geometries, but most organic molecules break apart at relatively modest temperatures. If a neural network potential is trained with an exponentially small number of high energy configurations, the predicted energies and forces at these geometries will be severely inaccurate which leads to pathological behavior, including fusion of nuclei and bond dissociation.

\indent Normal Mode Sampling (NMS) has been used successfully in generating NNMCs for atomic and molecular properties.\cite{smith2017ani,rupp2015machine}. These approaches begin with an optimized geometry, and then use a harmonic model of the local PES with reference forces to ensure even energy sampling. For small distortions this is nearly optimal sampling, but large distortions look very different than reactive paths because the linear normal modes are poor approximations of the curvy motions executed by atoms. See Figure S1 in the supplementary information for examples. Further, the normal mode coordinates depend on the hessian matrix, whose computation scales at least quadratically with system size, making NMS prohibitively costly for large training molecules and systems. Most state-of-the-art neural network model chemistries only depend on energy, force and perhaps some electron density information, and so the added overhead of computing the Hessian is a significant limitation. 

\indent In this paper we consider metadynamics (MetaMD)\cite{laio2002escaping,valsson2016enhancing,barducci2011metadynamics,laio2005assessing,bussi2006equilibrium,raiteri2006efficient} as an efficient way to sample molecular geometries. MetaMD implements bias potentials against previously visited geometries based on a collective variable. It is used in MD of large molecules to encourage sampling of rare events, typically in conjunction with hand-chosen collective variables which encourage the desired large scale motion. In this work we simply use the matrix of atomic distances: $D_{ij} = 1/|\vec{r_{ij}}|$ as the collective variable. Note that since the distance matrix is an important ingredient in most neural network descriptors, this bias essentially guarantees that sampling will generate dissimilar descriptor vectors to previous samples. All atoms are democratically forced to new environments in a way which respects the invariances of molecular energies. Unlike high temperature MD, MetaMD does not simply combust molecules, it samples near-equilibrium for as long as it takes to saturate unique distance matrices. 

\begin{figure}[t!]
    \centering
    \includegraphics[width=0.45\textwidth]{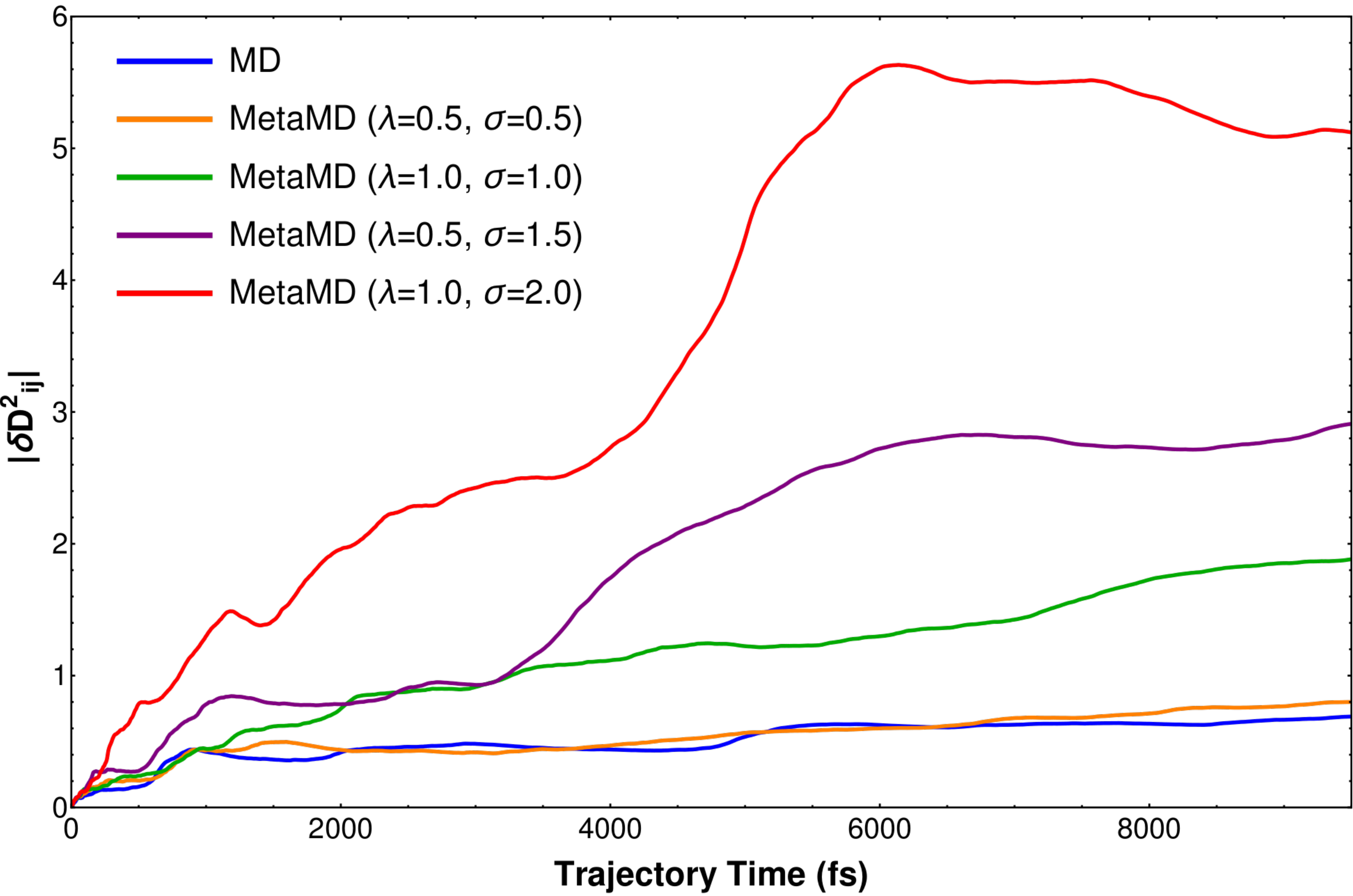}
    \caption{Running expectation of the distance matrix variance with ordinary Andersen dynamics and MetaMD $\langle |(\delta D_{ij})^2|\rangle (t)$. Larger $\lambda$ and $\sigma$ values provide more bias against previous snapshots of the distance matrix.}
    \label{fig:var_distmat}
\end{figure}

\section{Methods}

\subsection{Formalism}
\indent We will describe the implementation of MetaMD sampling in our open-source TensorMol package\cite{TENSORMOL}, although the technique is easily implemented in any framework for machine learning since the forces are given trivially by algorithmic differentiation. TensorMol can be used to execute thermostatted MD using either an \emph{ab-initio} model or a sufficiently accurate NNMC as the atomic potential, $V(\vec{R})$. For a MetaMD propagation, we allocate space to store a large list of previous distance matrices $D_{ij}^\alpha$, which we accumulate every $\tau$ femtoseconds. If the molecule is especially large, this matrix of distances can be held in a sparse representation, and a neighbor list can be used to evaluate the force in the same fashion as our linear-scaling evaluation of symmetry functions.\cite{yao2017tensormol} We add to $V(R)$, a "bump" perturbation of the form,
\begin{align}
    V_\text{bump}(\vec{R}) = \sum_\alpha\lambda e^{-\sum_{ij}(D_{ij}(\vec{R}) -D_{ij}^\alpha )^2)/(2\sigma^2)}. 
\end{align}
Here, $\alpha$ sums over snapshots of geometries where a "bump" has been placed at a particular value of the collective variable, and $D_{ij}$ is the (possibly sparse) atomic distance matrix at a given point during the trajectory. There are three parameters in this accelerated exploration method: the bump height, $\lambda$, the bump width, $\sigma$, and the frequency of bumps, $\tau^{-1}$. When used with an ordinary NVE thermostat, the advantage of the scheme is that it explores new geometries near equilibrium without dissociating the molecule like high-temperature dynamics.

\indent In our provided software package, the MetaMD scheme is implemented in TensorFlow.\cite{tensorflow2015-whitepaper} Example code for the kernel and force are given in the supplementary information. This version of a bump function is ideal for descriptors which depend mostly on the distance matrix. It can be tuned to sample different types of molecular motion, by selectively neglecting blocks of $D_{ij}$. In fact really any descriptor can be used in place of $D_{ij}$ in this scheme. The bump responds to both small collective or large local changes to geometry. The enhanced sampling obtained can be seen from several features of the dynamics: from the statistical variance of geometry, by using dimensionality reduction techniques, and by testing the robustness of learned networks. We will discuss each of these in the results. 

\subsection{Sampling methods}
\indent We define the fitness of a PES sampling method practically, by comparing the accuracy of a network trained with those samples to data from another sampling method. The three schemes examined are the most common in the literature and the most reasonable. One might imagine generating samples by random cartesian or spherical perturbations to equilibrium coordinates, or random placements of atoms in space\cite{Korth:2009aa}. These very random methods produce very unique geometries, but cannot really be used for training accurate NNMCs because of multiple SCF solutions. They also sample high energy configurations far too often.  

\indent MD trajectories for nicotine and a ten water molecule cluster with time steps of 0.5 femtoseconds at a temperature of 600 Kelvin were propagated for sampling data. The cluster of water molecules was propagated for 4 picoseconds, and the nicotine molecule was propagated for 25 picoseconds. The MD was thermostatted with the Anderson thermostat, and the MD trajectory propagates using Velocity Verlet integration\cite{andersen1980molecular,verlet1967computer}.

\indent MetaMD was used with identical parameters as MD, but bias potentials were added every 10 femtoseconds according to the formalism above. MetaMD bumps height and width parameters will be explored as a part of our analysis and are given for each trajectory in the Results section.

\indent NMS follows the scheme as described by Smith and coworkers\cite{smith2017ani}. To generate representative samples, the set of normal mode coordinates is obtained at the desired level of theory, and a set of random displacement vectors are computed along a harmonic potential. The value of $k_{b}T$ was chosen to be 0.002 hartrees. The number of samples generated was chosen to match that of MD and MetaMD for data used to train neural networks.

\indent Sampling of geometries for MD and MetaMD was performed with a pretrained neural network, and sampled geometries were saved for subsequent \emph{ab-initio} calculations. All \emph{ab-initio} calculations were performed with the Q-Chem package\cite{shao2015advances} at the \textomega B97X-D/6-311g**\cite{chai2008long} level of theory.

\subsection{Neural network models}
\indent NNMCs are a rapidly evolving field, and there is already an inexhaustible diversity of input descriptors, network topologies, and fragmentation schemes which fall under this blanket term. Because we cannot examine every option, this paper uses the most common Behler-Parinello network topology\cite{behler2007generalized,behler2011atomcentered} with the ANI-1 modified symmetry functions\cite{smith2017ani} and a loss function that includes molecular energies as our `representative neural network model.' Optimization of network parameters was done using the ADAM optimizer in TensorMol.\cite{kingma2014adam}

\indent Many groups are investigating modifications to this popular strategy, but changes in the descriptor and network shape do not affect the need for large amounts of efficiently sampled training data. In this work, the network model is made up of three hidden layers that contain 500 neurons in each layer for each branch of the NN, and the network's activation function is modeled by the exponential linear unit, or ELU\cite{clevert2015fast}. We trained a total of 18 networks; one for each of the sampling methods as described above with increasing amounts of training data. Each network was trained for 2,000 epochs, and the network parameters which provided minimal test errors during training were saved as the final parameters of the NN. When evaluating the accuracy of our networks' energies, we calculated mean absolute errors (MAE) and root mean square errors (RMSE) of the networks' energies as they compare to \emph{ab-initio} calculations. MAE and RMSE results from all 18 networks are provided in Tables S1-5 of the supporting information.

\begin{figure}
    \centering
    \includegraphics[width=0.45\textwidth]{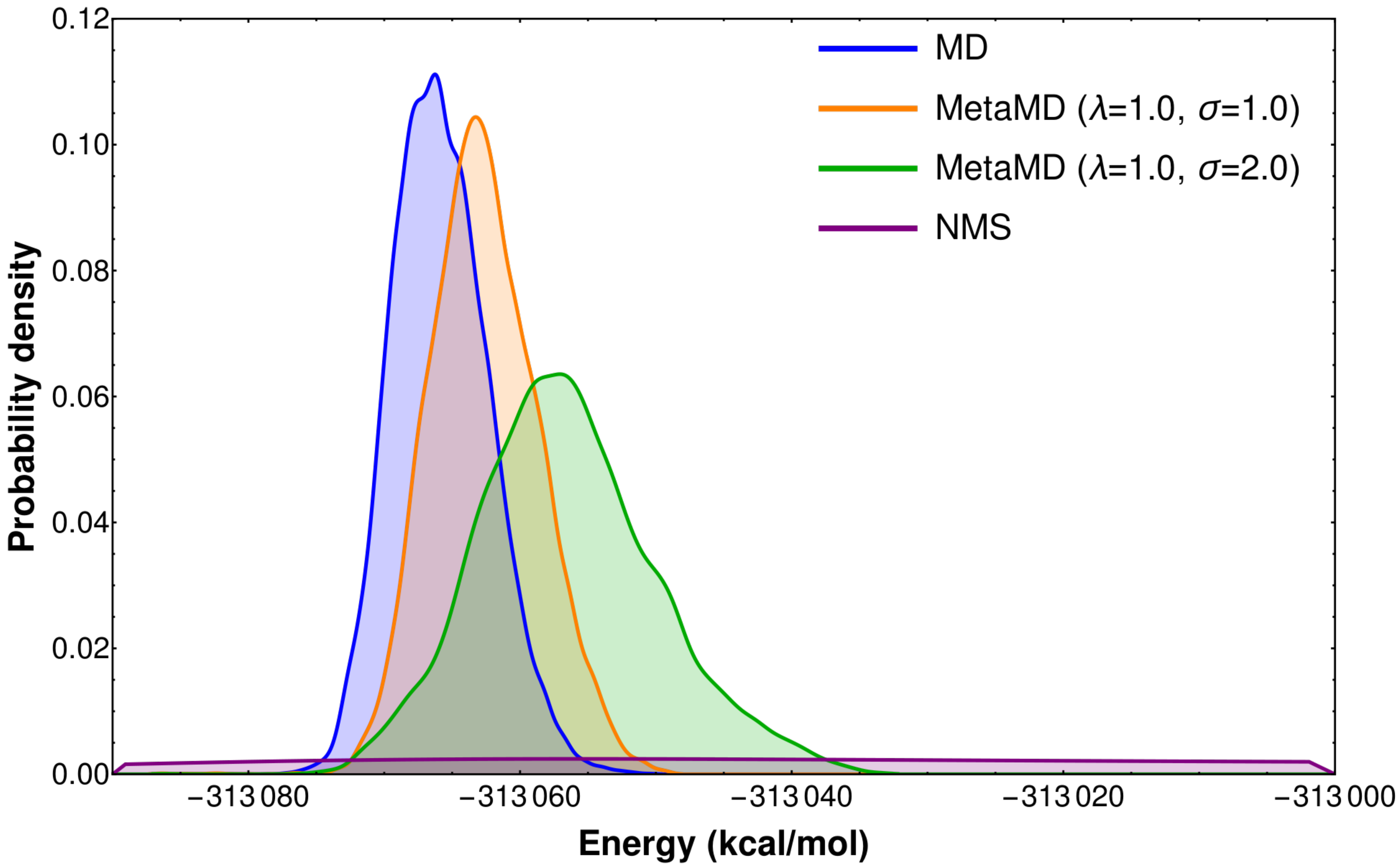}
    \caption{Histogram of energies from an MD trajectory (Blue), MetaMD with bump heights of 1.0 hartree in both cases and bump widths of 1.0 hartrees (orange) and 2.0 hartrees (green) and from NMS (purple).}
    \label{fig:energy_hist}
\end{figure}

\section{Results}
\indent The root mean square (RMS) variance of the distance matrix for different bump heights and widths used in MetaMD trajectories of nicotine, along with an MD trajectory for reference, is shown in Figure \ref{fig:var_distmat}. Increasing the bump height and width increases the variance of the distance matrix throughout the trajectory. We note that parameters of 1.0 hartrees for the height and 2.0 \AA for the width tends to give the best results for a single molecule without breaking bonds. Smaller values do not explore configuration space rapidly enough, while larger parameters can quickly break bonds. We note here that breaking of bonds can be desired if a reactive NNMC is desired, but for the purposes of this work we constrain ourselves to non-reactive NNMCs.

\indent Figure \ref{fig:energy_hist} shows the distribution in potential energies from an MD trajectory
using the Andersen thermostat\cite{andersen1980molecular} at 300 Kelvin, from two MetaMD trajectories with different bump parameters, and from NMS. Dynamics trajectories were run for 10 ps with a 0.5 fs time step. We note that NMS energies are distributed over a larger range than shown here, with a maximum sampled energy of -310,968 kcal/mol. For reference the largest energy sampled by any of the other three trajectories was -313,033 kcal/mol. From this histogram MD and MetaMD seem very similar, although we will show that their behavior for training networks is qualitatively different. We will also show that surprisingly although this histogram shows that the energies sampled by NMS and MetaMD are totally different, they produce networks of very similar accuracy. In short: a histogram of sampled energies is a pretty poor measure of how much chemical space is explored. To bring this to light it is better to look at the configurations of geometries themselves. We can use basic dimensionality reduction techniques to show that the space of configurations MD samples is small relative to Meta MD.  

\begin{figure}
    \centering
    \includegraphics[width=0.45\textwidth]{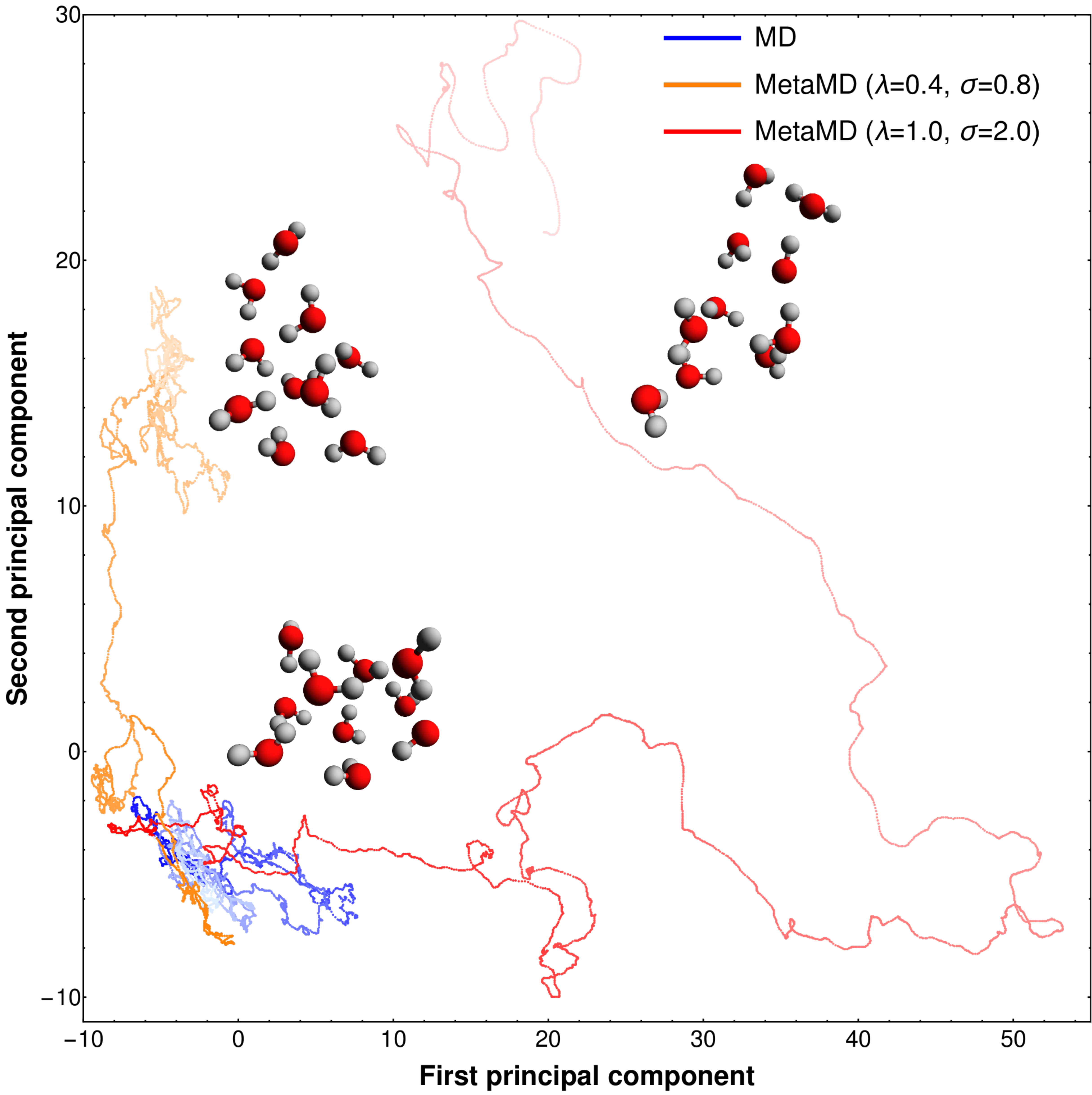}
    \caption{PCA of the distance matrix along a 4 picosecond trajectory with MD (Blue) and MetaMD with bump height (width) parameters of 0.4 (0.8) hartrees (Orange) and 1.0 (2.0) hartrees (Red) at a temperature of 300 K. All simulations starts with a cage-like minimal energy geometry. The color of each line fades as the trajectory time increases. Inserted figures are the final geometries after the 4 picosecond simulation.}
    \label{fig:pca}
\end{figure}

\indent Figure \ref{fig:pca} shows the first and second components from principal component analysis (PCA) on the distance matrix of a ten water molecule cluster along MD and MetaMD trajectories. The MD principal components do not change significantly over the course of the trajectory. Modest bump parameters for MetaMD increase the change to the second principal component over the course of the trajectory. In this linear measure of structural similarity, the amount of configuration space sampled behaves roughly linearly with the bump parameters.

\begin{figure}
    \centering
    \includegraphics[width=0.45\textwidth]{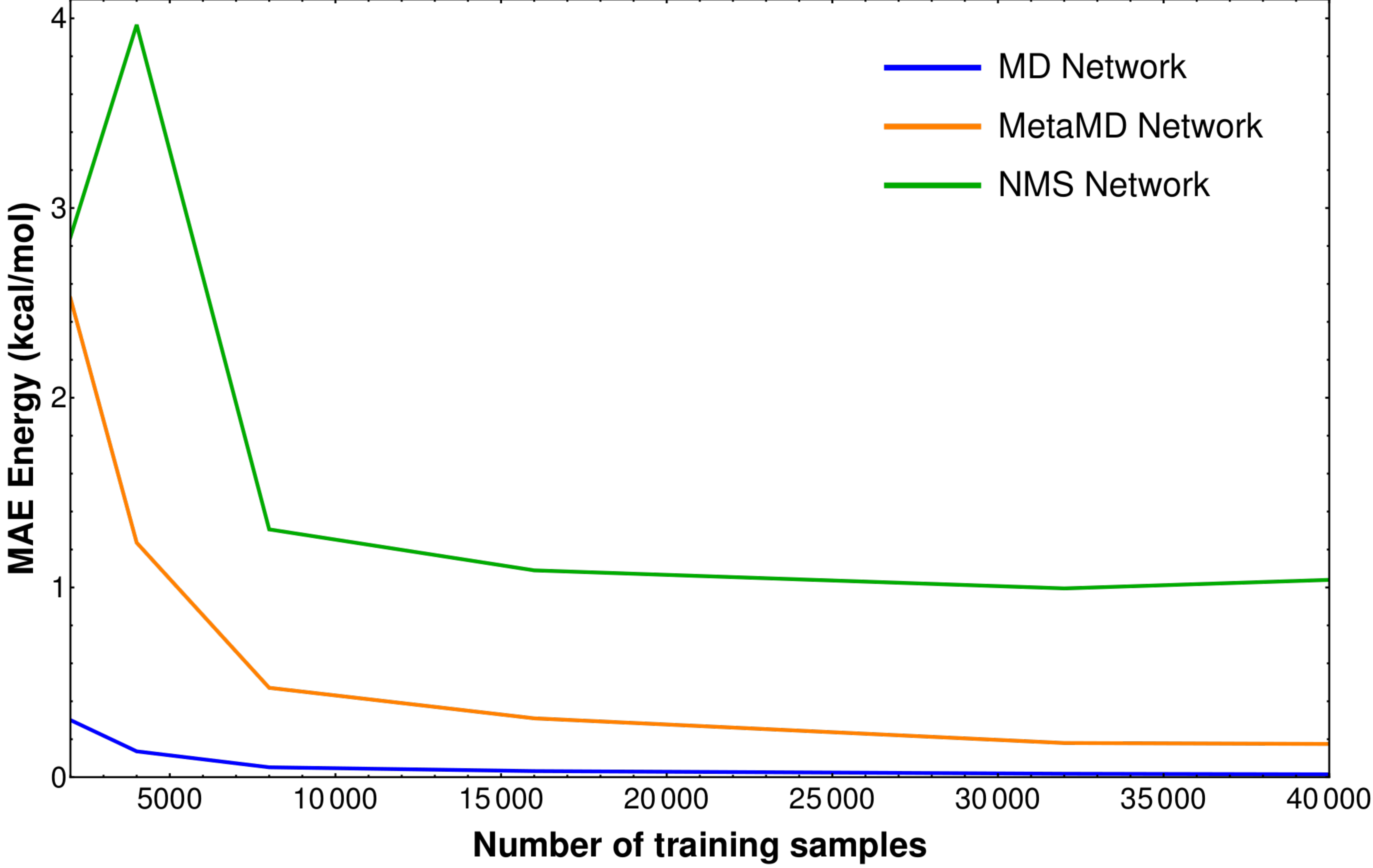}
    \caption{Mean absolute errors in the energy prediction evaluated on the same dataset as the number of training samples increases. AIMD converges with relatively training samples, while MetaMD and NMS both require more samples before converging.}
    \label{fig:selferror}
\end{figure}

\indent Next we look at the convergence of NNs with the number of training samples for each data generation method. By randomly sampling data from each method (after extracting some independent test data from each), we trained a series of NNs with increasing amounts of training data. Figure \ref{fig:selferror} shows the MAE relative to the parent DFT model chemistry as a function of the number of training samples. NNs trained with MD data converge to low errors with relatively few sample geometries provided for training, while NNs trained with MetaMD or NMS data require more training samples before the errors start to converge. \textbf{This figure is illustrative of an important cognitive pitfall which developers must avoid when producing NNMCs}. It is conventional for papers to select random test data from the same sort of sampling method used to generate training data. If an author were to do that with the data from Fig.\ref{fig:selferror}. they might erroneously conclude that MD is the best sampling technique, and the MD network is best because it has the lowest error on "independent test data". That is precisely the wrong conclusion. Most of the MD dataset is well represented by only a few sampled geometries ($\sim$ 8,000 geometries), because it is redundant. MetaMD and NMS networks both require more training data before converging ($\sim$ 32,000 geometries) because the networks are actually learning more of the PES. MetaMD and NMS both avoid repetitive sampling by forcing a larger and more even distribution of sampled geometries. To some extent the error histogram reveals the redundancy of the MD data, but it doesn't really allow much inference about the relative performance of MetaMD and NMS, as we will show when considering cross-sampling error. 

\begin{figure}
    \centering
    \includegraphics[width=0.45\textwidth]{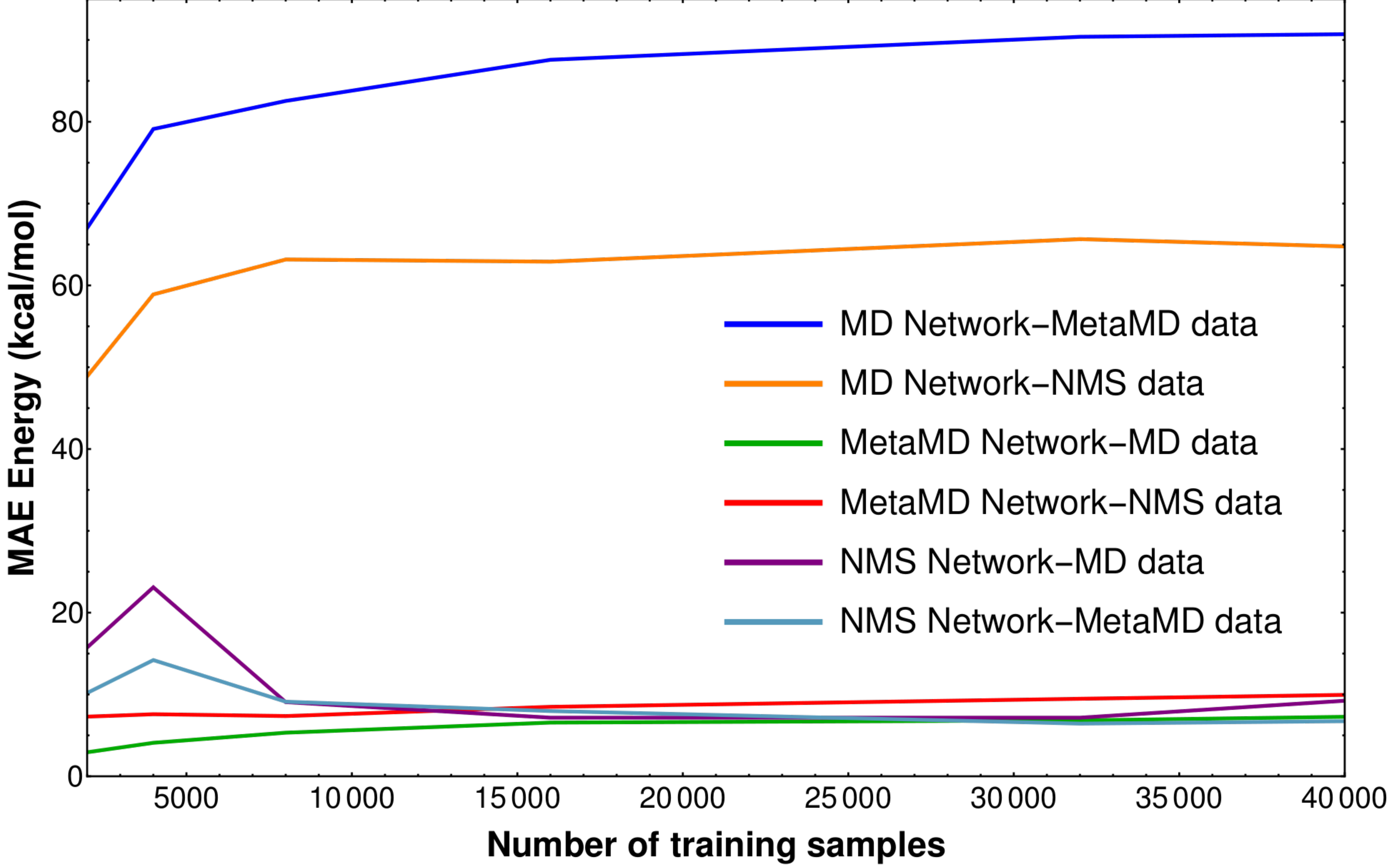}
    \caption{Mean absolute errors in the energy prediction cross evaluated on other dataset generation methods as the number of training samples increases. Labels are listed as training data-evaluation data. AIMD networks rarely see samples similar to geometries from MetaMD or NMS, while MetaMD and NMS networks both train on a larger variance of geometry samples, thus generalizing better across other data generation methods.}
    \label{fig:crosserror}
\end{figure}

\indent The whole picture becomes clearer when we look at the MAE of these NNs cross-evaluated on data from other generation methods. MD networks in particular do not generalize to MetaMD or NMS data with large MAE as shown in Figure \ref{fig:crosserror}. The MD network actually becomes \emph{less general} when trained on more MD data. This is a reversal of what most people would expect, and shows that "big data" is a small-player in the success of NNMCs unless the sampling is cleverly chosen or the test data is strongly correlated with the training set. MetaMD and NMS networks, however, provide similar accuracy on both datasets not included in their training data, with MAEs nearly and order of magnitude better than MD cross-evaluation errors. Given the natural similarity between MD and MetaMD, it is not surprising that MetaMD generalizes better to MD data than NMS both at small and large numbers of samples. It is also easy to rationalize the flat generalization of MetaMD to NMS data based off the histogram of energies sampled. The majority of the MetaMD error comes from very high energy samples which are rare in the MetaMD training with the amount of steps chosen. However, an advantage of the MetaMD approach is that after some linear number of additional steps, this region would eventually be sampled without intervention. 

\begin{figure}
    \centering
    \includegraphics[width=0.45\textwidth]{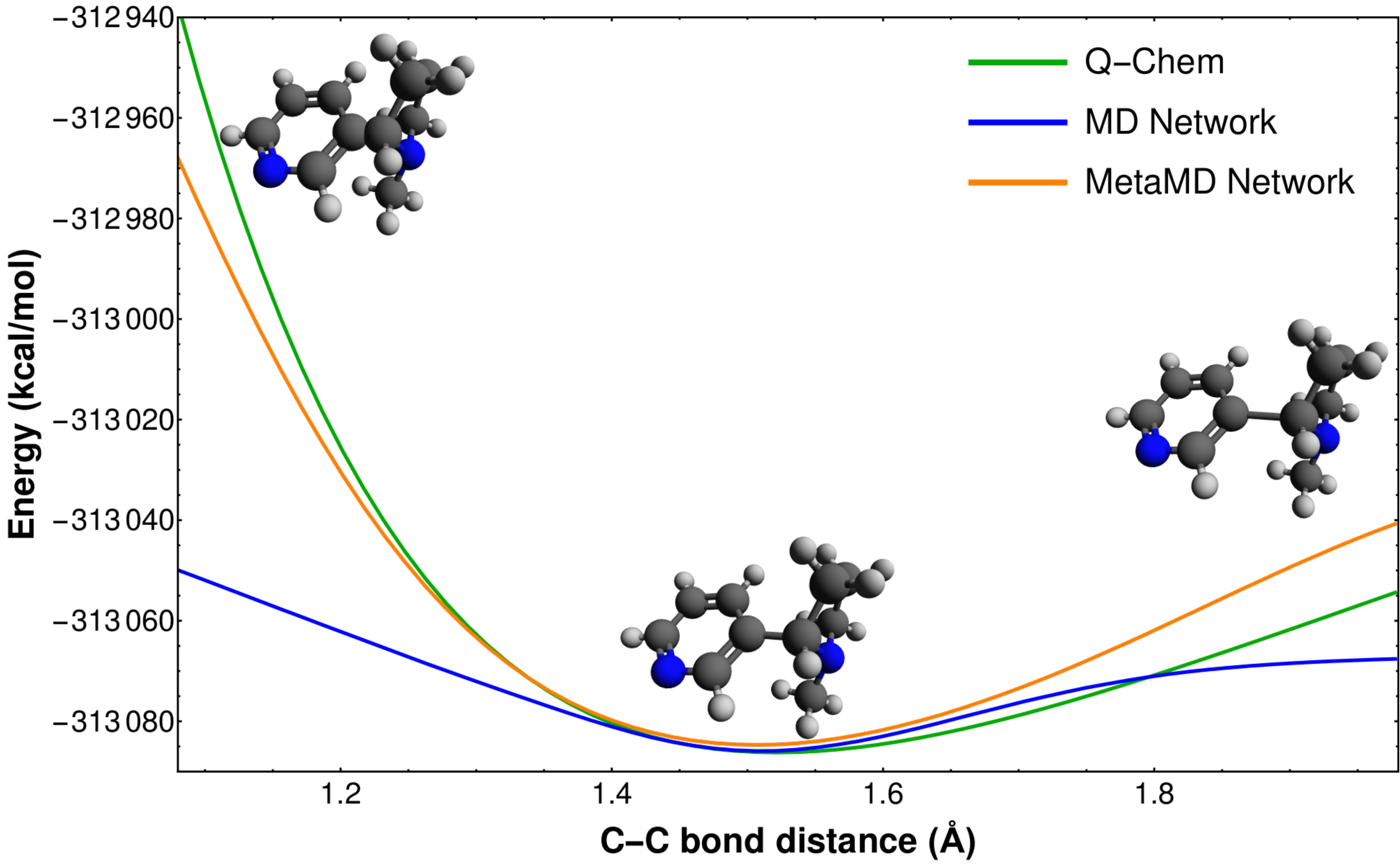}
    \caption{Potential energy curve for stretching and shrinking of the C-C bond connecting the rings of nicotine. Potential energies calculated from Q-Chem provided as a reference. The potential from a MetaMD network shows the desired sharp increase of the potential energy as the carbons become increasingly closer. The potential from a AIMD networks shows an increase in the potential energy, but does not rise as sharply as Q-Chem or the MetaMD network.}
    \label{fig:stretchcurve}
\end{figure}

\indent Large errors in the PES far from equilibrium can lead to catastrophic results when used to propagate dynamics. A network trained purely on MD data leads to such failures. Figure \ref{fig:stretchcurve} plots the potential energy of nicotine in an MD with an initial stretching of the carbon-carbon bond connecting the two ring structures. Both MD and MetaMD networks correctly predict the bonding minimum at 1.52 \r{A}. The MD network does not capture the steep inner potential as the carbon-carbon bond becomes short, while the MetaMD network correctly predicts this behavior. MD lacks any bonding interaction outside 1.8\AA, making bond-breaking artificially easy. 

\begin{figure}
    \centering
    \includegraphics[width=0.45\textwidth]{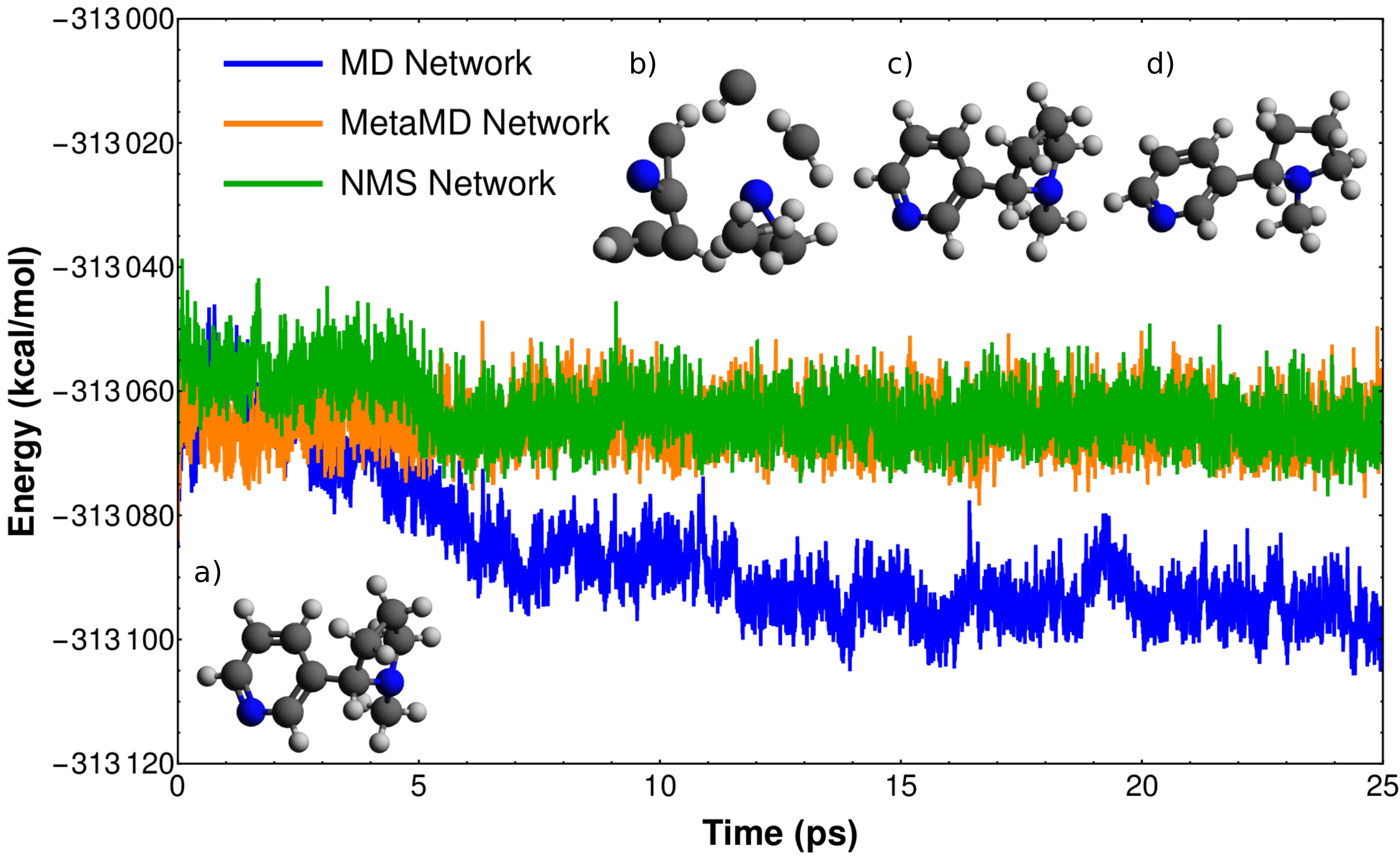}
    \caption{Energies from 25 picosecond dynamics trajectories with the Andersen thermostat at 300 Kelvin. The initial geometry (inset a) is the fully relaxed output from Q-Chem. The final geometries for propagating with the MD, MetaMD, and NMS networks are shown as insets b, c, and d respectively. The MD quickly blows the molecule apart, while the MetaMD and NMS networks are stable over the entire trajectory time.}
    \label{fig:mdtraj}
\end{figure}

\indent To exemplify how these errors occur in MD trajectories, we propagated 25 picosecond trajectories using networks trained on each data sampling method. The results are shown in Figure \ref{fig:mdtraj}. Variance of energies from MetaMD and NMS networks match well and remain close to the energy of the fully relaxed geometry. Energies from the MD network begin to drop well below the minimal geometry energy after 2-3 picoseconds. Examination of the final geometries from the dynamics trajectories show that the MetaMD and NMS network trajectories were stable across the entire trajectory time. The MD network, however, made significant distortions to the molecule before losing the bonding configuration all-together. This is due to artificially low bonding energies as seen in Figure \ref{fig:stretchcurve}.

\begin{figure}
    \centering
    \includegraphics[width=0.45\textwidth]{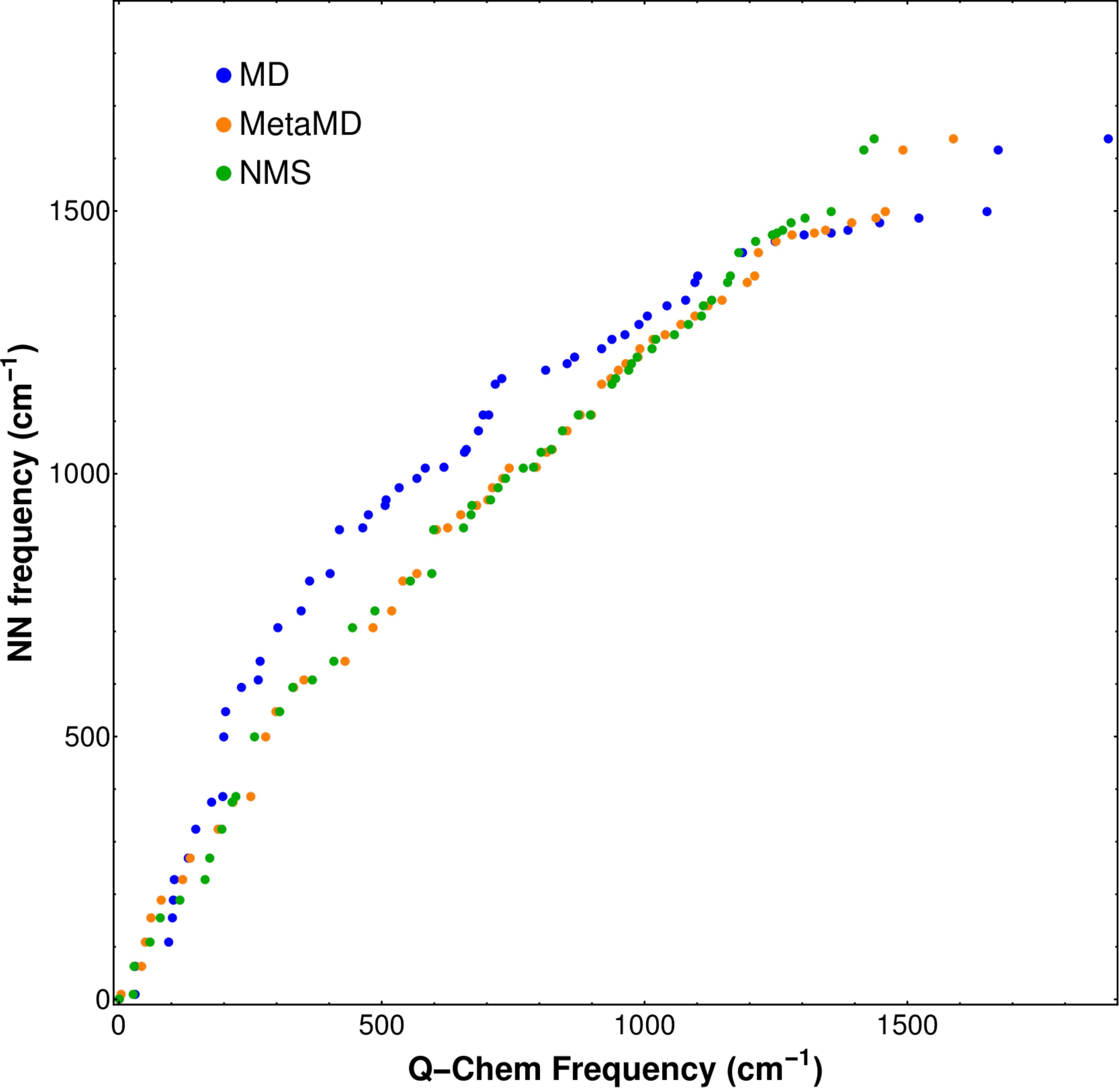}
    \caption{Predicted infrared frequencies from Q-Chem calculations vs those predicted from neural networks.}
    \label{fig:irfreq}
\end{figure}

It's clear that away from equilibrium the MD sampling network is riddled with unusable artifacts, so we examined a problem closer to equilibrium. Recent work has shown the ability of NNs to predict Infrared (IR) frequencies.\cite{gastegger2017machine,yao2017tensormol} We use our trained networks to predict the harmonic IR frequencies of nicotine and compare to those computed with Q-Chem as shown in Figure \ref{fig:irfreq}. To our surprise even harmonic force constants which only depend on small distortions away from equilibrium are predicted more poorly by the MD network than the MetaMD or NMS networks. These frequencies are based off a finite difference code, where most of the distortions are firmly within the energy distribution where MD samples most heavily, and the other two methods sample more sparsely. The highest energy configuration used during calculation of IR frequencies was -313081 kcal/mol. One possible explanation for this surprising behavior may be that the distant samples in MetaMD and NMS together with the requirement that the neural network PES is continuous, represent forces better in the loss function than additional small perturbations near equilibrium. The close agreement between MetaMD and NMS was a surprise to us, since NMS is somehow tailored to sample within a harmonic model. We note that all three networks were trained \emph{without} forces in the loss function for reasons of computational cost. Learning with forces is expensive, but enhances the accuracy of predicted IR spectra. 

\section{Discussion and Conclusions}
\indent     We have presented a comparison of several common methods used for sampling the PES when generating neural network training data, and presented a simple version of metadynamics which can be used to sample the PES in a naive black-box fashion with linear scaling cost. By comparing the performance of networks trained with these methods we showed that MD sampling is an unacceptably poor scheme. The MD network has deceptively small "test error" if test data is pulled from MD data, while actually the performance of the network becomes \emph{worse} in a general sense with additional training data.

\indent NMS which was recently used most notably by Smith and coworkers for the ANAKIN model chemistry\cite{isayev2015materials}, is shown to be an excellent sampling technique. However NMS requires the quadratic scaling computation of the hessian matrix in order to produce the sampled normal modes. MetaMD is linearly scaling, enforces sampling of novel geometries with novel embeddings, and is tunable such that desired regions of the PES can be sampled efficiently. It also is efficient in the sense that the generality of MetaMD vs. other sampling techniques saturates to a good value with a small number of samples. More even sampling of a range of energies could be achieved by further generalization of our MetaMD scheme, for example by choosing bump heights randomly from a uniform distribution. MetaMD can also be used to sample bond breaking events while producing reasonable geometries, although that is a matter for future work.

\indent    All three of the sampling methods discussed in this paper 'saturate' in the sense that additional samples do not improve generality after $\sim10^4$ samples of training data. Within-sampling independent test errors continue to decay exponentially, but the model is over-fitting to its sampling method's distribution of geometries. Further work on sampling algorithms should aim to push the ceiling of usable data towards larger numbers of samples. NNMCs are naturally over-complete, and practitioners must assume that they are operating within a regime of over-fitting to their training data. For this reason the method used to generate training samples must be considered at least as important as the definition of the model.

\indent We urge caution against a culture of testing NNMCs on benchmark datasets following the tradition of quantum chemistry \cite{jurevcka2006benchmark,curtiss1991gaussian,curtiss1999gaussian,curtiss2007gaussian,ramakrishnan2014quantum,smith2017ani} especially datasets near equilibrium. No measure of a NNMC to a single sampling technique contains quantitative information about general performance outside that sampling technique. Composite model chemistries such as G(N) theory\cite{jurevcka2006benchmark,curtiss1991gaussian,curtiss1999gaussian,curtiss2007gaussian} underfit and so in-sampling errors are a good measure of generality. NNMCs naturally overfit and so in sampling data is a poor measure of generality as shown here. Open source data and code\cite{smith2017ani,bereau2017non,smith2017anidata,TENSORMOL} is the best way to ensure that developments improve the general performance of these methodologies rather than a specific set or sampling method. Our code is available on github at http://github.com/jparkhill/TensorMol.

\bibliography{tensormol}

\end{document}